\documentclass[apex]{jjap3}
\usepackage{cite,overcite}

\title{Direct Comparison of Distant Optical Lattice Clocks at the $10^{-16}$ Uncertainty}
\author{Atsushi Yamaguchi$^{1,3}$, Miho Fujieda$^1$, Motohiro Kumagai$^1$, Hidekazu Hachisu$^{1,3}$, Shigeo Nagano$^1$, Ying Li$^1$, Tetsuya Ido$^{1,3}$\thanks{E-mail address: ido@nict.go.jp}, Tetsushi Takano$^{2,4}$, Masao Takamoto$^{2,4}$, and Hidetoshi Katori$^{2,4}$  }
\inst{$^1$ National Institute of Information and Communications Technology, 4-2-1 Nukui-kitamachi, Koganei, Tokyo 184-8795, Japan \\
$^{2}$ Department of Applied Physics, Graduate School of Engineering, The University of Tokyo, Bunkyo-ku, Tokyo 113-8656, Japan \\
$^3$ CREST, Japan Science and Technology Agency, 4-1-8 Honcho Kawaguchi, Saitama 332-0012, Japan \\
$^4$ Innovative Space-Time Project, ERATO, Japan Science and Technology Agency, Bunkyo-ku, Tokyo 113-8656, Japan.
}
\abst{Fiber-based remote comparison of $^{87}$Sr lattice clocks in 24 km distant laboratories is demonstrated. The instability of the comparison reaches $5\times10^{-16}$ over an averaging time of 1000 s, which is two orders of magnitude shorter than that of conventional satellite links and is limited by the instabilities of the optical clocks. By correcting the systematic shifts that are predominated by the differential gravitational redshift, the residual fractional difference is found to be $(1.0\pm7.3)\times10^{-16}$, confirming the coincidence between the two clocks. The accurate and speedy comparison of distant optical clocks paves the way for a future optical redefinition of the second.
}
\begin{document}
\maketitle

Over forty years, time and frequency have been stated by referencing the definition of the second, i.e., the Cs atomic clocks operating at microwave frequency. The clock frequencies are routinely monitored worldwide using satellite-based links\cite{Bauch}, which guarantee the coincidence of the frequencies generated in distant laboratories and constitute International Atomic Time. 
 On the other hand, recent dramatic advancement of optical atomic clocks' performance has urged Comit\'e International des Poids et Mesures (CIPM) to recommend four optical transitions of atoms or ions as secondary representations of the second\cite{BIPM}, which is a list of candidates for the future redefinition of the second. Among these transitions, the $^1S_0-^3P_0$ transition of $^{87}$Sr declared the least fractional uncertainty, $1\times10^{-15}$, based on the agreement obtained by three independent measurements\cite{JILA,SYRTE,NMIJ}. It is noteworthy that this optical clocks' uncertainty is actually constrained by that of the Cs clocks and relevant frequency links. The outstanding performance of optical clocks, therefore, will be truly revealed only if a direct optical comparison without interposing Cs clocks is established.

One of the challenges for a direct comparison is to establish an accurate and stable frequency link that faithfully transfers the optical clock signal to the remote site. Clearly, the best method would be to use an optical link instead of a satellite-based microwave link, as the higher carrier frequency improves the resolution and thus the stability and accuracy of the comparison. Coherent transfer of an optical signal over fiber lengths of 251 km was demonstrated with a residual instability below $2\times 10^{-16}$ at 1 s averaging time\cite{Newbury}. Using a 4-km-long optical fiber, a Sr lattice clock was referenced to a neutral Ca clock to evaluate systematic shifts\cite{SrYb}. Aiming towards a continental-scale frequency comparison, frequency transfers over 100-km-long fiber links between different laboratories were realized in Japan\cite{Musha}, France\cite{Jiang}, and Germany\cite{PTB146km}  with instability below $3\times10^{-15}$ at 1 s averaging time. However, direct comparison of the state-of-the-art optical clocks has not been attempted yet.

\begin{figure}
\begin{center}
\includegraphics[width=8cm]{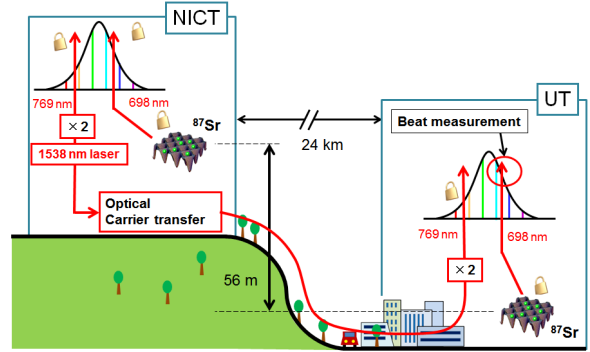}
\end{center}
\caption{Schematics of the optical frequency comparison between two distant lattice clocks via a telecommunication fiber network. The urban telecom fiber link operating at 1538 nm faithfully transfers a stable optical frequency from the Sr lattice clock at NICT (left) to UT (right). The differential frequency between two clocks can be precisely measured as a beat signal between the clock laser and the frequency comb at UT.}
\end{figure}

Figure 1 illustrates the experimental scheme of this study. The clock signal at the National Institute of Information and Communications Technology (NICT) is optically transferred to The University of Tokyo (UT) using a telecommunication fiber link. At NICT, the Ti:sapphire-based optical frequency comb\cite{Nagano} is phase-locked to the clock laser (698 nm). The telecom laser operating at the wavelength of 1538 nm, which is phase-locked to this optical frequency comb through its frequency-doubled light (769 nm), is transferred to UT through a phase-noise-cancelled\cite{MaLink} 60-km-long optical fiber. The transferred laser is frequency doubled and the Ti:sapphire-based optical frequency comb at UT is then phase-locked to it. The frequency difference between two Sr lattice clocks is monitored as a beat signal between the optical frequency comb and the clock laser at UT. According to the prior evaluation of the optical link including optical frequency combs, frequency doubling, and fiber noise cancelation, the fractional instability of the link can be estimated as $2\times 10^{-15}$ at the averaging time of 1 s and $7\times 10^{-17}$ at 1000 s. No frequency offset has been observed within the statistical uncertainty. The technical detail of the all-optical link is described elsewhere \cite{Fujieda}.

Two optical lattice clocks employing spin-polarized fermionic $^{87}$Sr atoms are developed separately at UT\cite{synchrocomparison,ltclk,Akatsuka}
 and NICT. An ensemble of roughly $10^4$ atoms is laser cooled\cite{muka} to $3\ \mu {\rm K}$ and loaded to the vertically oriented one-dimensional (1D) optical lattice potentials. After optically pumping to one of the stretched magnetic sublevels, the $5s^2\ {}^1S_0 \left(F=9/2, m_F =\pm 9/2\right)- 5s5p\  {}^3P_0 (F=9/2, m_F = \pm9/2)$ transition is probed using a clock laser propagating along the strong confinement axis of the lattice to suppress the Doppler and recoil shifts. By probing both sides of a lineshape at its full width at half maximum, the deviation of the clock laser frequency from the atomic resonance is detected and used for clock stabilization. At UT, the clock transition is observed with the Fourier-limited linewidth of 4 Hz for a 200 ms interrogation time. In a clock cycle time of 1.5 s, atoms are loaded into lattices and one side of one of the magnetic sublevels $(m_F = \pm9/2)$ is probed. At NICT, the clock transition is observed with the Fourier-limited linewidth of 20 Hz in a clock cycle time of 1.3 s.

\begin{table}
\caption{Systematic frequency corrections $\Delta$ and  uncertainties $\sigma$ for two $^{87}$Sr lattice clocks at UT and NICT, in units of Hz.}
\label{t1}
\begin{center}
\begin{tabular}{l|rr|rr}
\hline
Contributor & \multicolumn{2}{c}{UT} & \multicolumn{2}{c}{NICT}\\
& $\Delta$ (Hz) & $\sigma$ (Hz) & $\Delta$ (Hz) & $\sigma$ (Hz) \\
\hline
AC Stark - Lattice & 0.19 & 0.10 & 0.10 & 0.10 \\
AC Stark - Probe & 0.00 & 0.00 & 0.01 & 0.01 \\
Blackbody & 2.17 & 0.10 & 2.26 & 0.10 \\
2nd Zeeman & 1.24 & 0.10 & 0.23 & 0.10 \\
Gravitational shift & -0.95 & 0.09 & -3.57 & 0.05 \\
Collision & 0.00 & 0.10 & -0.04 & 0.12 \\
Servo error & 0.00 & 0.01 & 0.00 & 0.01 \\
\hline
Total & 2.65 & 0.22 & -1.01 & 0.22
\end{tabular}
\end{center}
\end{table}

 The systematic frequency corrections in each clock are independently evaluated as summarized in Table 1. The wavelength of the lattice laser is stabilized to the ``magic'' wavelength where the AC Stark shift for the ground and excited states becomes equal, leading to suppression of the differential scalar and tensor AC Stark shift to -0.19 (10) Hz at UT and -0.10 (10) Hz at NICT. The polarization direction of the linearly polarized lattice laser is parallel to the bias magnetic field. The higher-order AC Stark shift (hyperpolarizability) is negligibly small $(< 3\ {\rm mHz})$ in both systems\cite{SYRTEshift}. The first-order Zeeman shift and the vector AC Stark shift are suppressed by taking the center of transitions from stretched magnetic sublevels of spin-polarized atoms\cite{TakaJPSJ}. Due to the Pauli exclusion principle, the ensemble of ultracold spin-polarized fermions is also beneficial for suppressing the collisional shift to 0.00 (10) Hz at UT and 0.04 (12) Hz at NICT. The second-order Zeeman shift caused by bias magnetic fields of $230\ \mu {\rm T}$ and $99\ \mu {\rm T}$ is estimated to be -1.24 (10) Hz and -0.23 (10) Hz at UT and NICT, respectively\cite{SYRTEshift}. The elevation of the lattice clock at UT from Earth's geoid surface and correction of the corresponding gravitational shift are $20.37 \pm 2$ m and -0.95 (9) Hz, respectively, and $76.33 \pm 1 {\rm m}$ and -3.57 (5) Hz at NICT. The overall systematic frequency shift of the frequency difference,  $\nu_{\rm NICT} - \nu_{\rm UT}$, amounts to 3.66 (31) Hz.

 \begin{figure}
\begin{center}
\includegraphics[width=8cm]{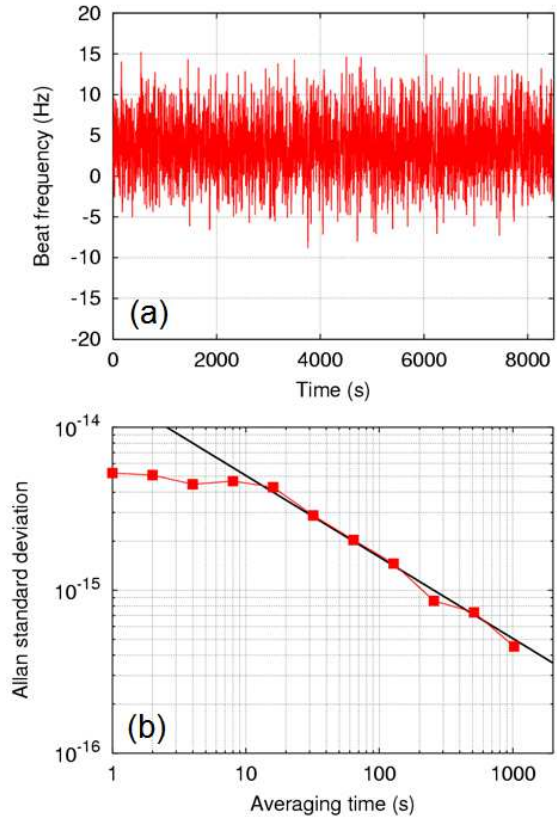}
\end{center}
\caption{Real-time observation of the frequency difference and stability between distant optical lattice clocks. (a) The frequency difference obtained by the beat measurement at UT is shown. The frequency difference caused by differential systematic shifts between the two clocks (3.66 Hz) is clearly detected. (b) The relative stability is shown as a function of averaging time. The obtained stability, $1.6\times 10^{-14} / \sqrt{\tau}$, is consistent with the Dick-effect-limited stability of the clock at NICT.
}
\end{figure}

 Figure 2(a) shows the time record of the frequency difference, $\nu_{\rm NICT} - \nu_{\rm UT}$. Thanks to the highly stable optical link, a Hz-level frequency difference between distant $^{87}{\rm Sr}$ optical lattice clocks is clearly visible over the time scale of minutes. The observed frequency difference is attributed to different systematic shifts between two clocks as listed in Table 1. It is noticeable that the largest contributor to the frequency difference is the gravitational shift of 2.62 Hz.

 The instability between the two clocks is measured to be $1.6\times 10^{-14}/ \sqrt{\tau}$  as shown in Fig. 2(b). It is noteworthy that the averaging time of 1000 s is sufficient to reach a fractional instability of $5\times 10^{-16}$, which indicates more than two orders of magnitude improvement over the remote comparison via the best current satellite-based microwave link\cite{Bauch}. The intrinsic noise of the clock laser and the dead time in the clock cycle cause aliasing noise that is referred to as the Dick effect\cite{Dicke}. The Dick-effect-limited instability is expected to be $6.0\times 10^{-15}$ for the clock at UT and $1.5\times 10^{-14}$ for the clock at NICT. These instabilities are consistent with the result shown in Fig. 2(b). This remote comparison system, therefore, allows us to investigate the relative instabilities of distant Sr lattice clocks that were previously masked by the instabilities of Cs clocks and relevant microwave links\cite{JILA,SYRTE,NMIJ,PTB}.

Frequency differences have been evaluated as summarized in Fig. 3, by taking eleven separate measurements over five weeks. In each measurement, we correct the corresponding systematic frequency shifts, which vary from day to day by less than 10 mHz due to small fluctuations in experimental conditions. The thin blue error bar indicates the systematic uncertainty of 0.31 Hz, except for the data on January 26th where the uncertainty is 0.49 Hz. The bold red error bar shows the standard error for each run that has measurement records in the range of 900 to 12000 s, from which the weighted mean of  $\nu_{\rm NICT} - \nu_{\rm UT}$ is calculated to be 0.04 Hz, as shown by the solid black line in Fig. 3. This result demonstrates that the two distant Sr lattice clocks generate the same frequency within the systematic uncertainty of 0.31 Hz ($7.3\times10^{-16}$ fractionally) for the 429 THz carrier frequency. This uncertainty is shown in Fig. 3 by dashed lines.

In summary, we have demonstrated for the first time the stringent and expeditious evaluation of distant optical clocks using optical fiber links, which significantly surpass previous frequency comparisons employing Cs clocks or satellite links. The uncertainty of the reproducibility identified here would be further reduced by rigorously managing the systematic shifts\cite{Swallows, SYRTEshift} and using less noisy fiber networks accordingly. Such endeavors will certainly push forward the optical redefinition of the second. The technique discussed here has a wide range of applications\cite{Review}.
Recently invented compact frequency combs based on microresonators\cite{comb} can be stabilized to the remote elaborate reference by the transfer technique demonstrated here. Fully referenced transportable optical atomic clocks will enable highly sensitive measurements in fieldwork. Synchronous frequency comparison between distant optical clocks\cite{synchrocomparison}, in which laser noise is canceled out as common noise, may uncover tiny temporal variations in the gravitational potential in real time, which might give new insights into geodesy.

\begin{figure}
\begin{center}
\includegraphics[width=8cm]{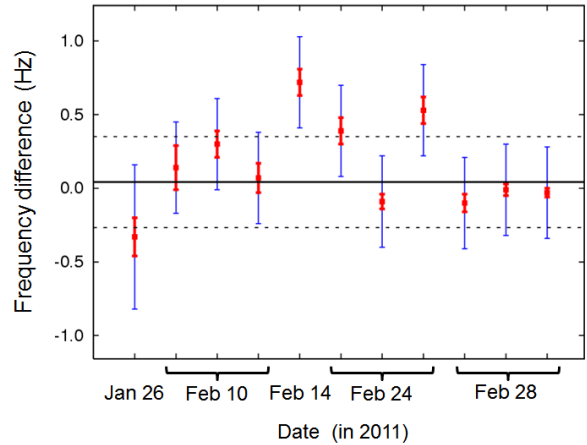}
\end{center}
\caption{Frequency difference between distant $^{87}{\rm Sr}$ optical lattice clocks. Systematic frequency shifts are corrected. Each point represents a data set obtained by the beat measurement of the duration 900-12000 s. The systematic uncertainty and standard error are shown as the blue thin and red bold error bars, respectively.
The weighted mean based on standard errors is 0.04 Hz and shown as a solid horizontal line. The black dashed lines express the overall systematic uncertainty shown in Table 1.
%
}
\end{figure}

\acknowledgments
We thank M. Hosokawa, Y. Koyama, N. Shiga, H. Ito, and K. Kido for their comments and experimental support. This research was supported in part by the Photon Frontier Network Program of MEXT and by JSPS through its FIRST program

\end{document}